\documentclass{WileyMSP-template}

\usepackage{graphicx}
\usepackage{dcolumn}
\usepackage{bm}
\usepackage{amssymb}
\usepackage{amsmath}
\usepackage{color}
\usepackage{subfig}
\usepackage{hyperref}
\usepackage{cite}

\begin{document}
\pagestyle{fancy}

\title{$\mathcal{PT}$-symmetric quantum Rabi model: Solutions and exceptional points}

\maketitle

\author{Jiong Li}
\author{Yi-Cheng Wang}
\author{Li-Wei Duan*}
\author{Qing-Hu Chen$^{\dagger}$}

\begin{affiliations}
J. Li, Y.-C. Wang\\
Address\\
Zhejiang Key Laboratory of Micro-Nano Quantum Chips and Quantum Control, School of Physics, Zhejiang University, Hangzhou 310027, China \\
L.-W. Duan*\\
Address\\
Department of Physics, Zhejiang Normal University, Jinhua 321004, China \\
Email Address:duanlw@zjnu.edu.cn  \\
Q.-H. Chen$^{\dagger}$\\
Address\\
Zhejiang Key Laboratory of Micro-Nano Quantum Chips and Quantum Control, School of Physics, Zhejiang University, Hangzhou 310027, China \\
Collaborative Innovation Center of Advanced Microstructures, Nanjing University, Nanjing 210093, China\\
Email Address: qhchen@zju.edu.cn 
\end{affiliations}

\vspace{1em}
\textit{This is the pre-peer reviewed version of the following article:J. Li, Y.-C. Wang, L.-W. Duan, Q.-H. Chen, $\mathcal{PT}$-Symmetric Quantum Rabi Model: Solutions and Exceptional Points. Adv Quantum Technol. 2025, 8, 2400609, which has been published in final form at https://doi.org/10.1002/qute.202400609. This article may be used for non-commercial purposes in accordance with Wiley Terms and Conditions for Use of Self-Archived Versions.}
\vspace{1em}


\begin{abstract}
The $\mathcal{PT}$-symmetric non-Hermitian quantum Rabi model (QRM) with imaginary coupling is solved using the Bogoliubov operators approach. A transcendental function responsible for the exact solutions is derived, with its zeros yielding the regular spectrum. We find two types of intersections: One is the exceptional point (EP), which is widely studied in the non-Hermitian system; another one is due to doubly degenerate states caused by the conserved QRM parity, which is well-known in the Hermitian QRM. These intersections are identified through this transcendental function. EPs emerge between pairs of adjacent excited energy levels, shifting toward lower coupling strengths as energy levels increase. The fidelity susceptibility diverges to negative infinity at the EPs, consistent with recent findings in non-Hermitian systems, while it diverges to positive infinity at the doubly degenerate points. The EPs are further confirmed by the vanishing c-product in the biorthogonal basis. All eigenstates are characterized by conserved energy and QRM parity. We conclude that the non-Hermitian QRM is integrable, analogous to its Hermitian counterpart.
\end{abstract}

\section{Introduction}

Non-Hermitian systems recently attracted considerable attention \cite{ashida_non-hermitian_2020}. In contrast to Hermitian systems, non-Hermitian systems allow energy exchange with the environment, resulting in Hamiltonians that typically produce complex spectra. Non-Hermitian Hamiltonians arise in various fields of physics, including cold atomic systems \cite{longhi_non-hermitian_2013, lee_heralded_2014, li_non-hermitian_2022}, superconducting vortex systems \cite{feinberg_non-hermitian_1999}, atomic and molecular physics \cite{moiseyev_non-hermitian_2011}, and surface hopping \cite{gao_non-hermitian_2017, edvardsson_sensitivity_2022}. Several theoretical approaches have been developed to explore their distinctive properties, such as the Feshbach projection \cite{ashida_non-hermitian_2020}, nonunitary conformal field theory \cite{mostafazadeh_pseudo-hermiticity_2002-1, bender_pt_2015, fring_exact_2017}, and biorthogonal quantum mechanics \cite{mostafazadeh_pseudo-hermiticity_2002, brody_biorthogonal_2014, tzeng_hunting_2021}.

Under certain conditions, a non-Hermitian Hamiltonian with parity-time ($\mathcal{PT}$) symmetry can exhibit an entirely real eigenvalue spectrum \cite{bender_real_1998, konotop_nonlinear_2016, el-ganainy_non-hermitian_2018}. For a $\mathcal{PT}$-symmetric Hamiltonian $H$ where $H \vert \phi \rangle = E \vert \phi \rangle$, the parity-time operator $\mathcal{PT}$ satisfies $\mathcal{PT} H \mathcal{TP} = H$, which leads to $\mathcal{PT} H \vert \phi \rangle = H \mathcal{PT} \vert \phi \rangle = E^{\ast} \mathcal{PT} \vert \phi \rangle$. Consequently, the eigenvalue spectrum is either complex-conjugate or entirely real when $\mathcal{PT}$ symmetry is preserved. This concept of $\mathcal{PT}$ symmetry has been effectively applied to systems with controlled dissipation \cite{ruter_observation_2010, feng_single-mode_2014}, and trapped ions \cite{wang_observation_2021}. Exceptional points (EPs), also known as branch points, mark the transition between symmetric and broken phases, where eigenvalues become complex. Near an EP, several intriguing phenomena have been observed, including band-merging \cite{li_exceptional_2023}, unidirectional invisibility \cite{ozdemir_paritytime_2019}, and fast self-pulsations \cite{miri_exceptional_2019}. These findings underscore the novel effects of EPs in non-Hermitian physics.

The quantum Rabi model (QRM) is the simplest model describing the light-matter interaction between a two-level system and a single-mode cavity \cite{rabi_process_1936, scully_zubairy_1997, braak_semi-classical_2016}. The QRM finds broad applications across various physical fields, including cavity and circuit quantum electrodynamics (QED) systems \cite{niemczyk_circuit_2010}, solid-state semiconductor systems \cite{forn-diaz_ultrastrong_2017, forn-diaz_ultrastrong_2019}, trapped ions \cite{leibfried_quantum_2003}, and quantum dots \cite{englund_controlling_2007, hennessy_quantum_2007}. The two-level system in the QRM serves as a fundamental qubit for implementing quantum simulations and computations \cite{cong_frequency-renormalized_2017, xie_quantum_2017}. The exact analytical solution for the QRM remained elusive until Braak presented a transcendental function, termed as the $G$-function, using the Bargmann space representation \cite{braak_integrability_2011}. Soon after, this $G$-function was derived in a more physical context using the Bogoliubov operator approach (BOA) \cite{chen_exact_2012}.

The non-Hermitian semi-classical Rabi model with $\mathcal{PT}$ symmetry has garnered significant attention in recent studies. Lee introduced a purely imaginary coupling constant, resulting in gain and loss dynamics within the two-level system \cite{lee_pt_2015}. Exact Floquet solutions have been identified for certain EPs in a time-periodic $\mathcal{PT}$-symmetric Rabi model, particularly under conditions of multiple-photon resonance \cite{xie_exceptional_2018}. The fully quantized Rabi model (QRM) can also exhibit $\mathcal{PT}$ symmetry and offer potential applications for open quantum systems \cite{cao_exact_2021}. When a purely imaginary bias is introduced in the QRM \cite{lu_pt_2023}, multiple EPs emerge as the light-matter coupling strength increases, contrasting with the single EP typically observed in most non-Hermitian system phase diagrams.

In this work, we extend the non-Hermitian semi-classical Rabi model to a fully quantized version by replacing the classical optical field with its quantum counterpart and investigating the non-Hermitian QRM with imaginary coupling constants. Imaginary coupling has been employed to describe open quantum systems, such as atomic vapors and optical waveguides \cite{peng_anti-paritytime_2016, zhang_dynamically_2019, xu_nonreciprocity_2020}. In the non-Hermitian QRM, imaginary coupling is achievable through dissipation-induced interactions incorporating nonlinearity and synthetic magnetism \cite{xu_nonreciprocal_2021, wang_dissipation-induced_2022, tian_interplay_2023, ying_robust_2024}. A feasible experimental nonreciprocal magnonic device operating at the single-magnon level has been proposed, where the magnon mode is coherently and dissipatively coupled to a superconducting qubit, producing a non-Hermitian Jaynes-Cummings model \cite{wang_dissipation-induced_2022, ying_robust_2024, ying_universal_2024}. Therefore, these systems could potentially implement the non-Hermitian QRM as the coupling strength increases.

The structure of this paper is organized as follows: Section II generalizes the $G$-function technique to the non-Hermitian QRM, providing an analytical derivation of the exact eigenvalue spectrum. Section III identifies two types of intersections: doubly degenerate points and exceptional points (EPs). These intersections are detected using the $G$-function technique and validated through fidelity susceptibility and the c-product. Finally, Section IV summarizes our findings and discusses the integrability of the non-Hermitian model.
	
\section{Solutions within Bogoliubov Transformation}
	
The Hamiltonian of the non-Hermitian QRM is expressed as
\begin{equation}
    H = - \frac{\Delta}{2} \sigma_{x} + \omega a^{\dagger}a + ig \left( a+a^{\dagger} \right) \sigma_{z}, \label{H_NHQRM}
\end{equation}
where $a$ and $a^{\dagger}$ denote the photon annihilation and creation operators of a single-mode cavity with frequency $\omega$. The term $ig$ represents the purely imaginary qubit-cavity coupling constant, $\Delta$ is the tunneling matrix element, and $\sigma_{x,z}$ are the Pauli matrices. For simplicity, we set $\omega = 1$ throughout this paper.
	
The parity operator is defined as
\begin{equation*}
    \mathcal{P} = \sigma_{x} \otimes \mathbf{1},
\end{equation*}
where $\mathbf{1}$ is the identity operator for the single-mode cavity. This differs from the corresponding QRM parity operator $\Pi = \sigma_{x} \otimes \exp \left[ i\pi a^{\dagger}a \right]$ in the Hermitian case. The time-reversal operator $\mathcal{T}$ performs complex conjugation, such that $\mathcal{T} \hat{x} \mathcal{T} = \hat{x}$, $\mathcal{T} \hat{p} \mathcal{T} = -\hat{p}$, where $\hat{x}$ and $\hat{p}$ represent the displacement and momentum operators respectively, implying that $\mathcal{T} a(a^{\dagger}) \mathcal{T} = a(a^{\dagger})$. Therefore,
\begin{eqnarray}
    \mathcal{PT} H \mathcal{TP} = \mathcal{P} \left( a^{\dagger}a - ig \left( a+a^{\dagger} \right) \sigma_{z} - \frac{\Delta}{2} \sigma_{x} \right) \mathcal{P} = a^{\dagger}a + ig \left( a+a^{\dagger} \right) \sigma_{z} - \frac{\Delta}{2} \sigma_{x} = H,
\end{eqnarray}
which confirms that this Hamiltonian is $\mathcal{PT}$-symmetric.
	
In contrast to the unitary transformation used in \cite{chen_exact_2012}, we apply a similarity transformation $D(ig) = \exp[ig \left(a^{\dagger} - a \right)]$, which satisfies the following conditions:
\begin{eqnarray}
    D(ig) a(a^{\dagger}) D(-ig) = a(a^{\dagger})-ig, \quad
    D(ig)D(-ig) = 1, \quad
    D(ig)^{\dagger}D(ig) = D(2ig) \neq 1. \nonumber
\end{eqnarray}
As a result, the Hamiltonian from Equation \eqref{H_NHQRM} is reformulated as
\begin{eqnarray}
   H_{s} = D(ig)HD(-ig)
   =
   \begin{bmatrix}
    a^{\dagger}a + g^{2} & -\frac{\Delta}{2} \\
    -\frac{\Delta}{2} & a^{\dagger}a - 2ig \left( a+a^{\dagger} \right) - 3g^{2}
   \end{bmatrix}
   , \label{H_QRM_S}
\end{eqnarray}
The general eigenfunction expansion for $H_{s}$ is proposed as
\begin{equation*}
    \vert \psi \rangle =
    \begin{bmatrix}
    \sum_{n=0}^{+\infty} i^{-n} \sqrt{n!} e_{n} \vert n \rangle \\
    \sum_{n=0}^{+\infty} i^{-n} \sqrt{n!} f_{n} \vert n \rangle
\end{bmatrix},
\end{equation*}
where $e_{n}$ and $f_{n}$ are the expansion coefficients, and $\{\vert n \rangle \}$ represent Fock states generated by $a^{\dagger}$ acting on the photon vacuum state. Using the Schr\"{o}dinger equation $H_{s} \vert \psi \rangle = E \vert \psi \rangle$ and projection onto $\vert n \rangle$, a recursive relation of $e_{n}$ and $f_{n}$ is derived:
\begin{subequations}
    \label{recur_QRM}
    \begin{equation}			
        e_{n}=\frac{ \frac{\Delta}{2} f_{n} }{ n+g^{2}-E },
        \label{recur_QRM_en}
    \end{equation}
    \begin{equation}
        f_{n+1} = \frac{ -\frac{\Delta}{2} e_{n} + \left( n-3g^{2}-E\right) f_{n} }{ 2g(n+1)} + \frac{f_{n-1}}{n+1}.  \label{recur_QRM_fn}
   \end{equation}
\end{subequations}
Without loss of generality, we set $f_{0}=1$. Consequently, all $f_{n \ge 1}$ and $e_{n \ge 0}$ can be determined from Equation \eqref{recur_QRM}.
	
Transforming back to the original Hamiltonian, we obtain
\begin{equation}
  \vert \Psi \rangle = D(-ig) \vert \psi \rangle
\end{equation}
Notably, the parity $\Pi = \sigma_{x} \otimes \exp [{i\pi a^{\dagger}a}]$ for the QRM also holds for the non-Hermitian QRM, i.e. $\vert \Psi \rangle \propto \Pi \vert \Psi \rangle$. Projecting onto the photon vacuum state $\vert 0 \rangle$, we obtain the $G$-function
\begin{eqnarray}
    G_{\pm} = \sum_{n=0}^{+\infty} \left( e_{n} \mp f_{n} \right) g^{n} = 0,
    \label{G_QRM}
\end{eqnarray}
whose zeros correspond to the eigenvalues of the Hamiltonian. The $\pm$ signs in the $G$-function (\ref{G_QRM}) represent positive and negative $\Pi$ parity, respectively, similar to the Hermitian case \cite{chen_exact_2012}.

\begin{figure}[tbp]
    \centering
    \subfloat{ \includegraphics[width=0.50\linewidth]{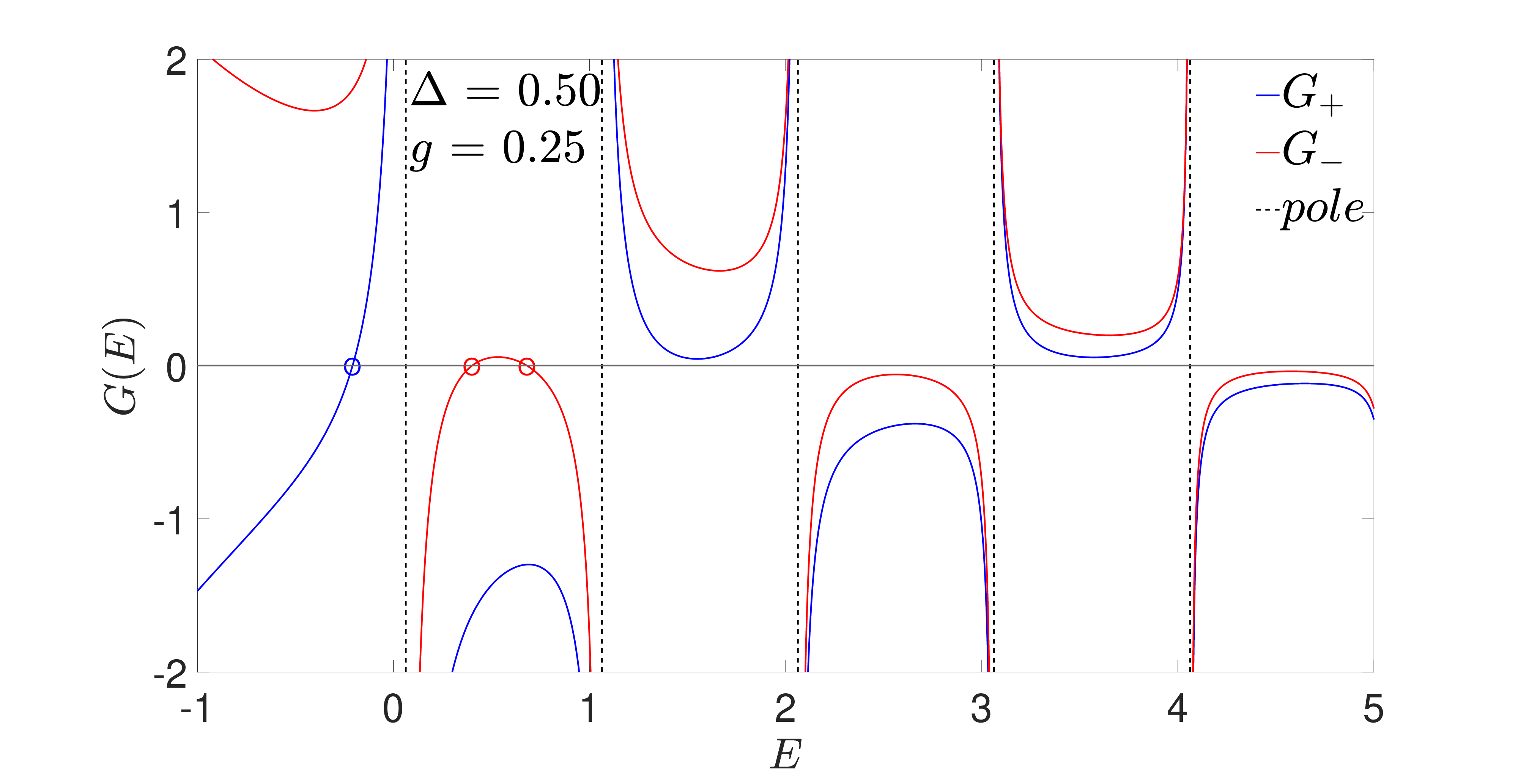} }
    \subfloat{ \includegraphics[width=0.50\linewidth]{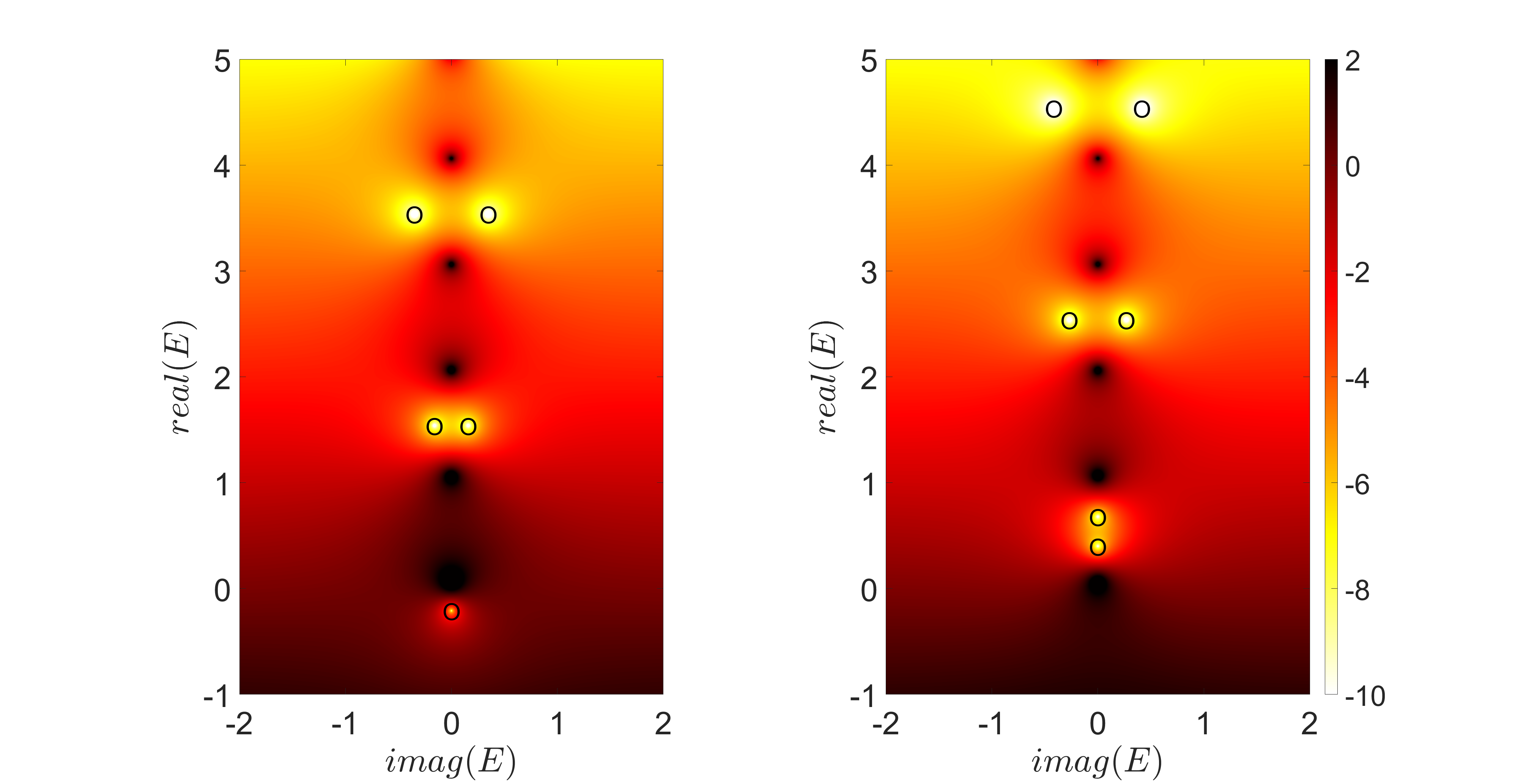}
    }
    \caption{ (colored online) The left panel displays the $G$-curves in the real $E$ regime. Blue (red) lines represents $G_{+}$ ( $G_{-}$ ) curves, and black dashed lines denote $E_{n}^{(pole)}$. The right panel gives $\ln |G_{+}|^{2}$ (left) and $\ln |G_{-}|^{2}$ (right) in the complex eigenvalue plane, respectively. Opened circles denote the zeros. $\Delta =0.50$ and $g=0.25$. }
    \label{G_fig}
\end{figure}

When $E$ is real, both $e_n$ and $f_n$ in Equation \eqref{recur_QRM} are also real, resulting in a real $G$-function. As shown in the upper panel of \textbf{Figure \ref{G_fig}}, real eigenenergies corresponding to the $\mathcal{PT}$-symmetric phase are determined by the zeros of the real $G$-function. However, as $E$ increases beyond a certain threshold, the real $G$-function no longer has real zeros. This absence of real zeros signifies broken $\mathcal{PT}$ symmetry, where real eigenenergies $E$ no longer exist. At this point, complex solutions for $E$ and the complex $G$-function must be considered. In this case, the zeros of the complex $G$-function (i.e., $\operatorname{Re}(G) = \operatorname{Im}(G)=0$) yield complex eigenvalues. Both real and complex zeros of the $G$-function are marked with open circles in the lower panel of Figure \ref{G_fig}.

\section{Two Types of Intersections}
 
\subsection{Doubly degenerate points}
\begin{figure}[tbp]
    \centering
    \includegraphics[width=0.8\linewidth]{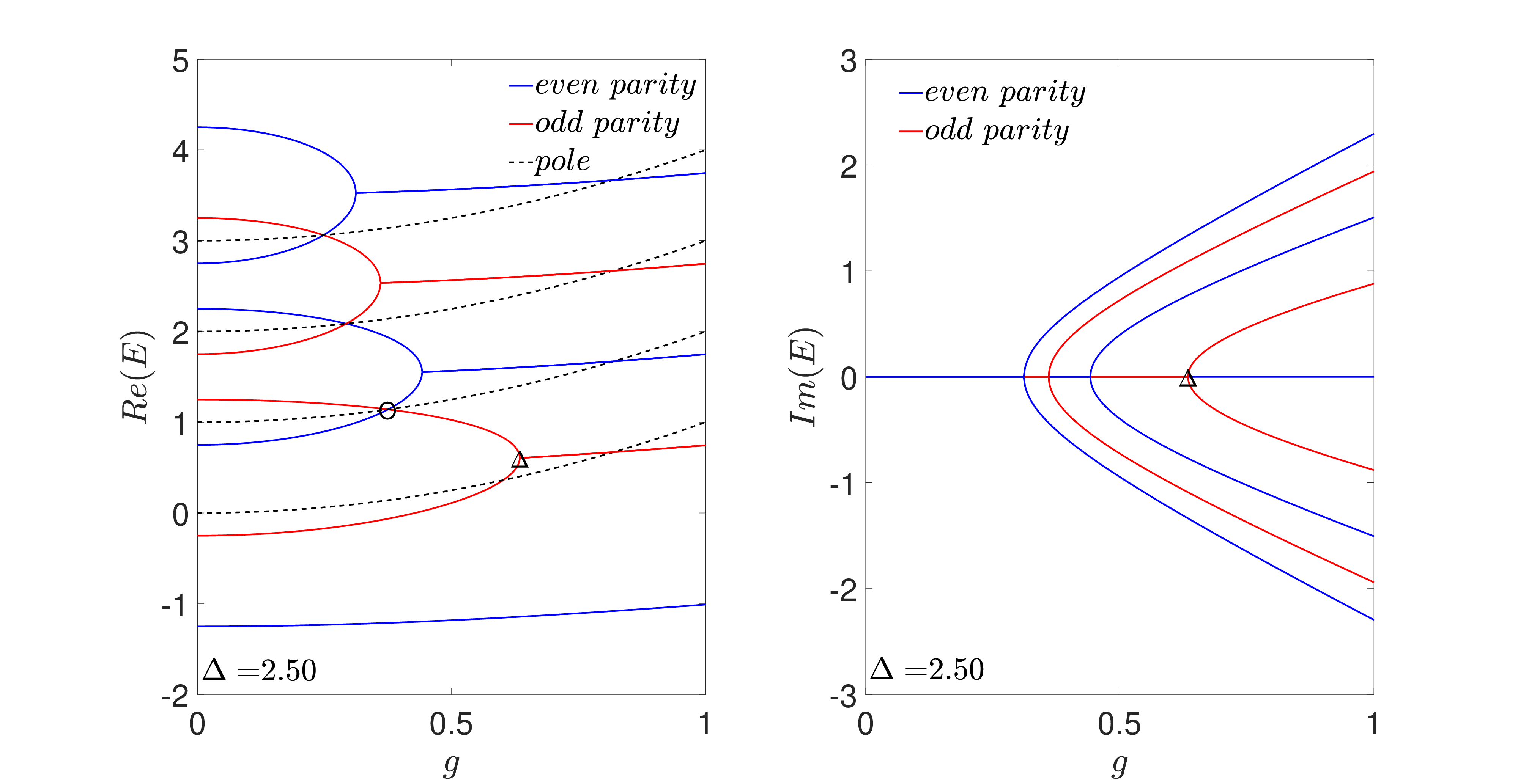}
    \caption{ (colored online) The left (right) panel displays the real (imaginary) part of the eigenvalue spectrum as a function of the coupling strength $g$. Blue (red) lines indicate levels within even (odd) $\Pi$ parity levels as determined by numerical diagonalization. Dashed lines correspond to $E_{n}^{(pole)}$. $\Delta = 2.50$. }
    \label{spectrum_d2.50_fig}
\end{figure}

The parity $\Pi$ is conserved when the corresponding energy level is non-degenerate. To identify the first type of intersection, namely doubly degenerate points, special attention should be paid to the pole lines, represented by black dashed lines in the upper panel of Figure \ref{G_fig}, and located at
\begin{equation}
    E_{n}^{(pole)} = n+g^{2}.
\end{equation}
The $G$-function diverges at these poles, which can be demonstrated using Equation \eqref{recur_QRM_en}. This pole structure closely relates to doubly degenerate eigenstates, a phenomenon well-studied in the Hermitian QRM \cite{braak_integrability_2011, chen_exact_2012, xie_double_2021}.  When $E = E_{n}^{(pole)}$, $e_{n}$ is well-defined only if
\begin{eqnarray}
    f_{n} = \frac{ - \frac{\Delta}{2} e_{n-1} + (n-1-3g^{2}-E_{n}^{(pole)}) f_{n-1} }{2ng} + \frac{f_{n-2}}{n} = 0. \nonumber
\end{eqnarray}
By iterating Equation \eqref{recur_QRM_fn}, the coupling strength $g_{n}$ can be determined, where the doubly degenerate state on the $n$th pole line occurs. For the first doubly degenerate state ($n=1$), the crossing of the second and third excited states in the energy spectrum occurs at 
\begin{eqnarray} 
    g_{1} = \frac{1}{2} \sqrt{ \frac{ \Delta^{2} }{4} - 1 }, \quad E_{1}= \frac{1}{4} \left( \frac{ \Delta^{2} }{4} + 3 \right).
\end{eqnarray}
The existence of a doubly degenerate state requires $\Delta \geq 2$. Adjacent levels with opposite $\Pi$ parities cross at the pole line, with the second and third levels intersecting precisely at $g=g_{1}$ and $E=E_{1}$, as circled in \textbf{Figure \ref{spectrum_d2.50_fig}}.
	
Thus, doubly degenerate eigenstates arise in the $\mathcal{PT}$-symmetric phase under specific conditions. In general, EPs can also be observed in Figure \ref{spectrum_d2.50_fig} where two real eigenvalues coalesce into a single value. This phenomenon will be further discussed in the next subsection.
	
\subsection{Exceptional points}
	
For a pair of complex-conjugate eigenvalues $E$ and $E^{\ast}$, Equation \eqref{recur_QRM_en} and \eqref{recur_QRM_fn} yield $e_{n}(E^{\ast}) = e_{n}^{\ast}(E)$ and $f_{n}(E^{\ast}) = f_{n}^{\ast}(E)$, resulting in $G_{\pm}(E^{\ast}) = G_{\pm}^{\ast}(E)$. Therefore, if $E$ is a solution of $G_{+}=0$ ($G_{-}=0$), then $E^{\ast}$ must also satisfy $G_{+}=0$ ($G_{-}=0$). As shown in the right panel of Figure \ref{G_fig}, the complex zeros of the $G$-function are symmetrically positioned about the $\operatorname{Im}(E)=0$ line, indicating complex-conjugate eigenvalues. As discussed above, a pair of complex-conjugate eigenvalues are determined simultaneously by the zeros of either $G_{+}$ or $G_{-}$, thus sharing the same parity.
	
\begin{figure}[tbp]
    \centering
    \includegraphics[width=0.8\linewidth]{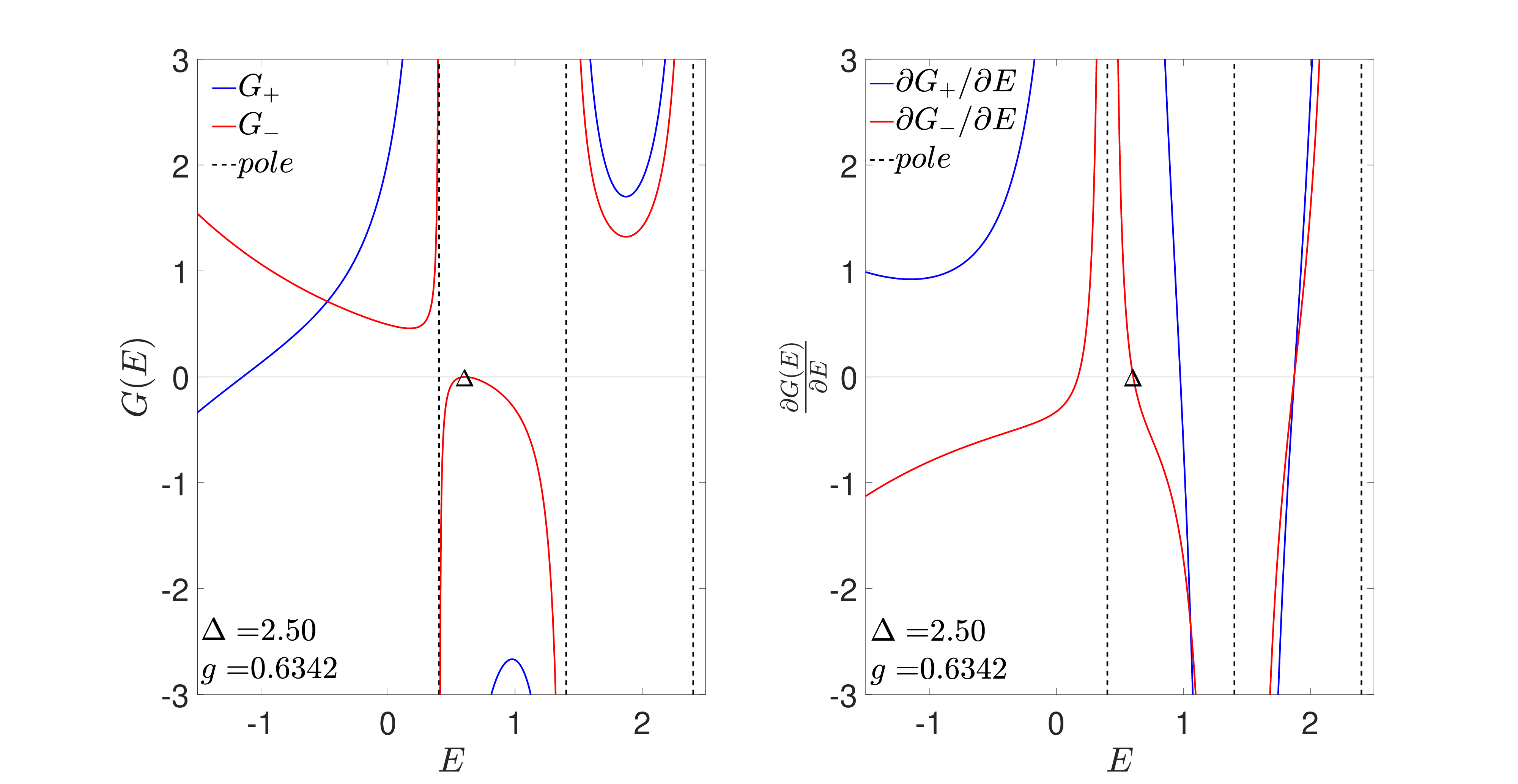}
    \caption{ (colored online) G-curves (left panel) and their first-order derivatives (right panel) with respect to $E$ are shown in the real $E$ regime at $\Delta=2.50$ and $g=0.6324$, matching the parameters for the EP marked by a triangle in Figure \ref{spectrum_d2.50_fig}. Blue (red) lines mark curves corresponding to even (odd) $\Pi$ parity, and black dashed lines denote $E_{n}^{(pole)}$. }
    \label{G_EP_fig}
\end{figure}

An EP indicates that the $G$-function has a single zero between two adjacent poles. As shown in \textbf{Figure \ref{G_EP_fig}}, the EP marked by a triangle in Figure \ref{spectrum_d2.50_fig} occurs when both the $G$-function and its first-order derivative with respect to $E$, $\partial G_{\pm }/\partial E$, simultaneously reach zero. Moreover, the zeros of the real $G$-function vanish at higher $E$ values, signaling $\mathcal{PT}$ symmetry broken.
	
This model provides an explanation for the coalescence of two eigenvalues and eigenvectors at EPs. At an EP, the two eigenvalues merge into a single value derived from the single $G$-function, resulting in identical eigenstates $\vert \Psi \rangle$, with the same coefficients $e_{n}, f_{n}$. While the doubly degenerate states cannot be obtained from the zeros of the $G$-function, there are two distinct states with opposite $\Pi$ parities at the crossing points.

\subsection{Distinguish between two types of intersections}
 
\begin{figure}[tbp]
    \centering
    \includegraphics[width=0.8\linewidth]{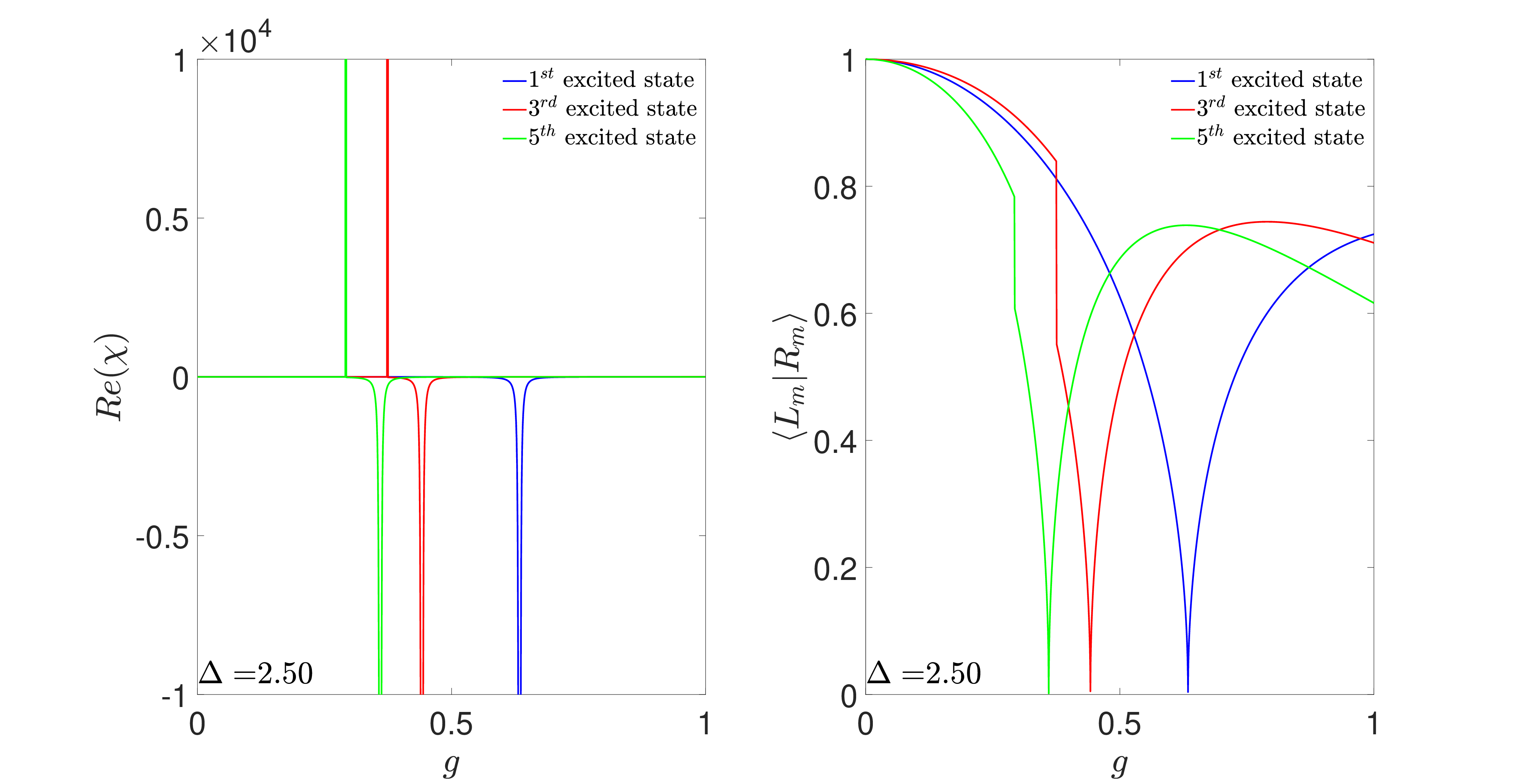}
    \caption{ (coloerd online) The real part of fidelity susceptibility (left panel) and c-product $\langle L \vert R \rangle$ (right panel)  as functions of the coupling strength $g$. The blue, red, and green lines correspond to the first, third, and fifth excited states at $\Delta =2.50$, respectively. }
    \label{fidelitysus_fig}
\end{figure}

Fidelity, a common measure for estimating the similarity between two quantum states, can be generalized as
\begin{equation}
    \mathcal{F} = \langle L(\lambda) \vert R(\lambda + \epsilon) \rangle \langle L(\lambda + \epsilon) \vert R(\lambda) \rangle,
\end{equation}
where $\vert L \rangle $ and $\vert R \rangle$ denote the bra and ket of the biorthogonal basis, respectively; $\lambda$ is a tunable parameter, and $\epsilon$ represents a minor disturbance \cite{tzeng_hunting_2021}. EPs can be identified by calculating the fidelity susceptibility,
\begin{equation}
  \chi = \frac{ 1 - \mathcal{F} }{ \epsilon^{2} },
  \label{fidelity_sus}
\end{equation}
The real part of $\chi(g)$, represented as $\operatorname{Re}(\chi(g))$, diverges to negative infinity as $g$ approaches an EP. This form reduces to the fidelity susceptibility of Hermitian systems and has been used to detect phase transitions. As illustrated in the left panel of \textbf{Figure \ref{fidelitysus_fig}}, $\operatorname{Re}(\chi(g))$ effectively identifies and distinguishes EPs from doubly degenerate points. Notably, $\operatorname{Re}(\chi(g))$ becomes extremely large and negative around EPs, whereas it is extremely large and positive near doubly degenerate points. In practical calculations using both the $G$-function technique and numerical diagonalization, $\operatorname{Re}(\chi(g))$ cannot reach true infinity due to Hilbert space truncation. Increasing the truncated space enhances the peak, suggesting the true divergence of the true fidelity susceptibility in both cases. In this sense, the truncated photon number serves as an effective system size.

The above EPs can be further corroborated by the self-orthogonal eigenvectors of the non-Hermitian Hamiltonian \cite{moiseyev_non-hermitian_2011}. By imposing the normalization condition $\langle L \vert L \rangle = \langle R \vert R \rangle = 1$, we calculate the so-called c-product $\langle L \vert R \rangle$, which represents the overlap of the bra and ket in the biorthogonal basis. As illustrated in the right panel of Figure \ref{fidelitysus_fig}, the c-product reaches zero at EPs, a result of the self-orthogonality property at these branching points.  Additionally, $\langle L \vert R \rangle$ exhibits a sharp jump at doubly degenerate points due to the parity switch in these eigenstates at such intersections. Therefore, the c-product provides a reliable method for distinguishing both types of intersections.
	
\section{Summary and Discussions}
	
In this work, we derived the $G$-function for the non-Hermitian QRM using the Bogoliuvbov operators approach, whose zeros determine the eigenvalue spectrum. In particular, within the $\mathcal{PT}$-symmetric regime, the $G$-function is defined in real variable space, enabling the identification of exceptional points (EPs) when both the $G$-function and its derivative with respect to $E$ are zero. Numerous EPs are located between pairs of adjacent eigenvalue levels, resulting in an infinite number of EPs. Furthermore, as the eigenvalue level increases, the EPs shift to lower coupling strengths. The pole structure of the $G$-function indicates the positions of doubly degenerate eigenvalues, similar to the Hermitian QRM. Significantly, we employ the newly developed fidelity susceptibility for non-Hermitian systems \cite{tzeng_hunting_2021} to distinguish EPs from doubly degenerate crossing points. The fidelity susceptibility diverges to negative infinity near $\mathcal{PT}$-symmetry-broken EPs, consistent with recent observations in other non-Hermitian systems, and diverges to positive infinity at doubly degenerate crossing points. Additionally, these EPs are confirmed through the c-product, highlighting the self-orthogonality characteristic at these branching points.
	
Finally, we discuss the integrability of the non-Hermitian QRMs, according to Braak's criterion of quantum integrability in the Hermitian QRM \cite{braak_integrability_2011}. Braak’s criterion states that a system is integrable if its eigenstates can be uniquely labeled by $f = f_{1} + f_{2}$ quantum numbers, where $f_{1}$ and $f_{2}$ represent the discrete (two-level) and continuous (photon) degrees of freedom, respectively. The non-Hermitian QRM discussed here also possesses $Z_{2}$ symmetry and the same number of symmetries (conserved energy and parity) as the Hermitian QRM. Even at EPs, where two eigenvalues coalesce within the same parity, the single resulting eigenstate can be labeled by conserved energy. Therefore, we may conclude that the non-Hermitian QRM is also integrable.

\medskip
\textbf{Acknowledgements} \par 
This work is supported in part by the National Natural Science Foundation of China (Grants No. 11834005 and No. 12305032).

\medskip

\end{document}